\documentclass[twocolumn,aps,prb,amsmath,amssymb,longbibliography]{revtex4-1}
\usepackage{xcolor}
\usepackage{graphicx}
\usepackage{bm}

\begin{document}

\title{Electronic Properties,
Screening and Efficient Carrier Transport in NaSbS$_2$}

\author{Jifeng Sun}
\author{David J. Singh}

\affiliation{Department of Physics and Astronomy, University of Missouri,
Columbia, MO 65211-7010, USA}

\email{singhdj@missouri.edu}

\date{\today}

\begin{abstract}
NaSbS$_2$ is a semiconductor that was recently shown to have
remarkable efficacy as a solar absorber indicating efficient
charge collection even in defected material. We report first principles
calculations of properties that show (1) an indirect gap only slightly
smaller than the direct gap, which may impede recombination of
photoexcited carriers, (2) highly anisotropic electronic and
optical properties reflecting a layered crystal structure, (3)
a pushed up valence band maximum due to repulsion from the Sb $5s$
states and (4) cross-gap hybridization between the S $p$ derived
valence bands and the Sb $5p$ states. This latter feature leads to
enhanced Born effective charges that can provide local screening
and therefore defect tolerance.
These features are discussed in relation to the performance of the
compound as a semiconductor with efficient charge collection.
\end{abstract}


\maketitle

\section{Introduction}

NaSbS$_2$ is a little studied semiconductor that was recently shown
to be remarkably effective as a solar absorber material, \cite{rahayu}
and has also been mentioned as a potential thermoelectric material based
on computational screening.  \cite{bhattacharya}
Remarkably,
an efficiency of 3.18\% was obtained in
the first report using NaSbS$_2$ nanparticles in a dye sensitized solar cell.
\cite{rahayu}
This is comparable to the efficiency of early
cells of similar type made using organometallic halide perovskites.
\cite{kojima}
Those materials have since proven to be a revolutionary advance in
photovoltaics,
but suffer from issues with long term stability and
the fact that they contain Pb, which is undesirable.
\cite{lee,giustino}
NaSbS$_2$ contains only environmentally friendly low cost elements,
and this very promising early experimental result, and the theoretical
results below suggest that it may as well represent a revolutionary new
material for solar photovoltaic applications.

From a valence point of view one might regard the stoichiometry as
derived from PbS by splitting of the divalent Pb site into
monovalent Na and trivalent Sb.
In this way one might anticipate that NaSbS$_2$ would be a semiconductor
and that the band gap may be higher than that of PbS if normal
trends are followed due to the splitting of the cation site.
However, at ambient temperature the crystal structure deviates strongly
the rocksalt structure of PbS, as discussed below.
More significantly, the presence of Na suggests a propensity for
defects, e.g. Na off-stoichiometry.
Good charge collection requires a high carrier drift length,
which typically occurs in high quality defect free material, and so
the high performance of NaSbS$_2$ as a solar absorber is surprising.
However, several soft lattice solar materials
have been discovered, most notably CH$_3$NH$_3$PbI$_3$, \cite{lee}
where defects
do not seem to play the same detrimental role as in more traditional
materials such as CdTe. \cite{stranks}
Here we report first principles calculations aimed at understanding the
properties of this compound, especially in relation to its
use as a solar absorber.

\section{Structure and Methods}

The present calculations were performed within density functional
theory (DFT). The electronic structure and optical properties were
calculated using the general potential linearized augmented planewave (LAPW)
method, \cite{singh-book}
as implemented in the WIEN2k code. \cite{wien2k}
The total energy calculations and
relaxation of the atomic coordinates was done using the Perdew, Burke
and Ernzerhof (PBE)
generalized gradient approximation (GGA). \cite{pbe}
For this, relativity was treated at a scalar relativistic level for the 
valence states. The core states were treated relativistically.
LAPW sphere radii of $R$=2.2 bohr were used for all elements,
along with a planewave sector basis cutoff determined by, $RK_{max}$=9
(here $R$ is the radius of the smallest LAPW sphere, i.e. 2.2 bohr, and
$K_{max}$ is the planewave cutoff).
Local orbitals were added to the basis for the S $s$, Na $s$ and $p$,
and Sb $d$ semicore states.

Spin-orbit was included for the electronic and optical properties.
The band gap is important for these, and accordingly these
calculations were done using the modified Becke-Johnson (mBJ) potential of
Tran and Blaha. \cite{mbj}
This functional gives band gaps in good accord with experiment for a wide
variety of simple semiconductors and insulators
and also appears to give reliable band shapes and optical properties,
although at least in certain semiconductors
the band masses are more similar to those obtained in standard
density functional calculations than those from many body calculations.
\cite{mbj,koller-mbj,singh-mbj,kim-mbj,singh-halide}
Calculation of the transport function for conductivity was done using
the BoltzTraP code. \cite{boltztrap}
Optical properties were calculated based on electric dipole transitions
in the independent particle approximation as implemented in the WIEN2k code.
The Born effective charges and the dielectric
tensor were calculated using the density functional perturbation theory
(DFPT) with the PBE functional as implemented in the VASP code \cite{vasp}
(note that calculation of the dielectric tensor cannot be done with
the mBJ potential, since it is not an energy functional and therefore
cannot be used to evaluate lattice response).

The structure of NaSbS$_2$ has been refined into two different monoclinic
groups, $C2/c$ (\# 15) \cite{olivier} and $C2/m$ (\# 12). \cite{volk}
In addition there is a report
of a triclinic structure $P\bar{1}$ (\# 2). \cite{kanischeva}
These are all centrosymmetric structures.
The deviation from a cubic structure was discussed in
terms of lone pair activity of Sb. \cite{olivier}

\section{Results and Discussion}

We did calculations for the three reported structures, in each case using the
lattice parameters from the diffraction experiments, and then relaxing
the atomic positions subject to the spacegroup symmetry. We find that the
energy for the $C2/m$ structure is 0.127 eV per formula unit (f.u.)
above the energy of the $P\bar{1}$ structure. The $C2/c$ structure is
0.005 eV/f.u. higher than the $P\bar{1}$ structure.

We did further calculations to address the issue of the ground state.
Specifically, we did full relaxations, including both lattice parameters,
angles and internal coordinates. For this purpose we used the VASP code
with three different density functionals, specifically the local density
approximation (LDA), the PBE GGA \cite{pbe} and the PBEsol GGA. \cite{pbesol}
We further did full relaxations of
both the $C2/c$ and the $P\bar{1}$ structures using VASP
with LDA, PBE, and PBEsol \cite{pbesol} functionals.
These functionals differ in equilibrium volumes for solids. Generally,
at increased volume lattices tend to distort more strongly,
as was noted in the case of PbTiO$_3$, an oxide ferroelectric with a ground
state particularly sensitive to volume.
\cite{wucohen}
We find that the LDA underestimates the unit cell
volume of NaSbS$_2$, yielding 175 \AA$^3$, in comparison with the
experimental volume of 192 \AA$^3$ at 300 K. The PBE functional
yields 197 \AA$^3$, while PBEsol yields 184 \AA$^3$. The LDA
predicts a monoclinic ground state, while the fully relaxed triclinic
structure is $<$ 2 meV/atom lower in energy for both PBE and PBEsol.
Considering the very small energy, and the limitations of DFT calculations we
conclude that the ground state is monoclinic
$C2/c$ or possibly $P\bar{1}$
with an extremely small triclinic distortion from this monoclinic structure.
The other monoclinic structure,
$C2/m$, is, however, clearly not a feasible structure.

\begin{figure}
\includegraphics[width=\columnwidth,angle=0]{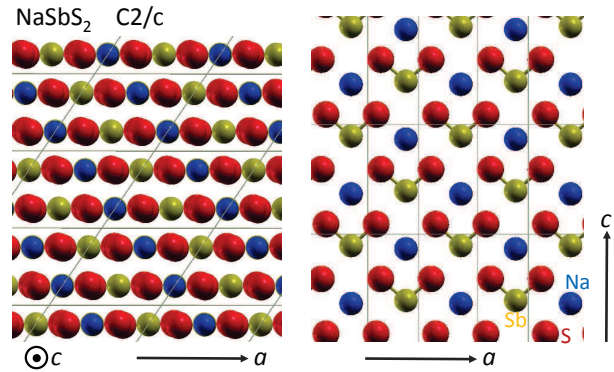}
\caption{Monoclinic structure of NaSbS$_2$, showing S as large red
spheres, Sb as gold and Na as blue. The left panel shows the layering,
while the right panel shows a single layer (note that there are two
layers with opposite orientation of the
S-Sb-S units  per cell.) The short Sb-S bonds in the S-Sb-S units are
shown by pipes.}
\label{fig-struct}
\end{figure}

We calculated properties for both the
$C2/c$ and $P\bar{1}$ structures, but find very little difference. For
example, the band gap for the triclinic structure is 1.21 eV compared to 
1.22 eV for the monoclinic.
This is not surprising since the two structures are very similar, though
not identical.
Internal coordinates for the two
structures are given in Tables \ref{tab-mono} and \ref{tab-tri}.
As seen, the bond valences \cite{brown}
are close to their nominal values indicative
of an ionic structure, Na$^+$Sb$^{3+}$S$_2^{2-}$,
although the Sb value of 2.85 is slightly smaller
perhaps indicative of some degree of covalency.

\begin{table}
\caption{Calculated atomic
positions and bond valence sums for monoclinic
NaSbS$_2$,
spacegroup 15, $C2/c$, $a$=8.232 \AA, $b$=6.836 \AA,
$c$=8.252 \AA, $\gamma$=55.72$^\circ$.
These lattice vectors are from experiment. ``b.v." denotes
the bond-valence sum. The fractional atomic
coordinates are in terms of the lattice
vectors and were determined with the PBE functional.}
\begin{tabular}{lcccc}
\hline
~~~~~~~~ & $x$ & $y$ & $z$ & ~~~b.v.~~ \\
\hline
Na & 0.0000 & 0.7500 & 0.1340 & 1.06 \\
Sb & 0.0000 & 0.7500 & 0.6051 & 2.85 \\
S  & 0.2239 & 0.7606 & 0.4088 & 2.02 \\
\hline
\end{tabular}
\label{tab-mono}
\end{table}

\begin{table}
\caption{Calculated atomic
positions and bond valence sums for triclinic
NaSbS$_2$,
spacegroup 2, $P\bar{1}$,
$a$=5.825 \AA, $b$=5.828 \AA, $c$=6.833 \AA,
$\alpha$=113.46$^\circ$, $\beta$=113.48$^\circ$, $\gamma$=90.07$^\circ$.
The atomic positions were determined using the PBE functional
and the lattice parameters are from experiment.}
\begin{tabular}{lcccc}
\hline
~~~~~~~~ & $x$ & $y$ & $z$ & ~~~b.v.~~ \\
\hline
Na & 0.3658 & 0.6332 & 0.2501 & 1.06 \\
Sb & 0.8947 & 0.1049 & 0.2499 & 2.84 \\
S  & 0.1330 & 0.3156 & 0.7393 & 2.02 \\
S  & 0.3154 & 0.1320 & 0.2390 & 2.02 \\
\hline
\end{tabular}
\label{tab-tri}
\end{table}

In the following, we discuss properties
of the monoclinic $C2/c$ structure for simplicity.
The structure is depicted in Fig. \ref{fig-struct}.
As seen, it is a layered structure, with two layers per unit cell.
The layers have composition NaSbS$_2$, with all atoms coplanar in
the monoclinic structure and very nearly coplanar in the triclinic structure.
A single layer is depicted in the right panel of Fig. \ref{fig-struct}.
The layers show distinct short bonds between Sb and two 
neighboring S in the same layer leading
to apparent S-Sb-S units. These units would lead to a strong ferroelectricity
in plane with polarization along the $c$-direction, except that the two
layers per cell have opposite orientation so that the polarizations of the
individual sheets cancel. The formation of S-Sb-S units leads to a dimerization
of the S in the layers. The near neighbor Sb-S distance is 2.486 \AA, while
the S-Sb-S angle is 98.6$^\circ$. The chains of S-Sb-S units run along the 
$a$-direction. Thus the structure has anisotropic layers in the $a$-$c$
plane.
In the following discussion of optical and
electronic properties we use a orthogonal coordinate system where
$x$ is along the $a$-axis, $z$ is along the $c$-axis and $y$ is perpendicular
to these. The tensor properties show an $xy$ component due to the
monoclinicity.

\begin{figure}
\includegraphics[width=\columnwidth,angle=0]{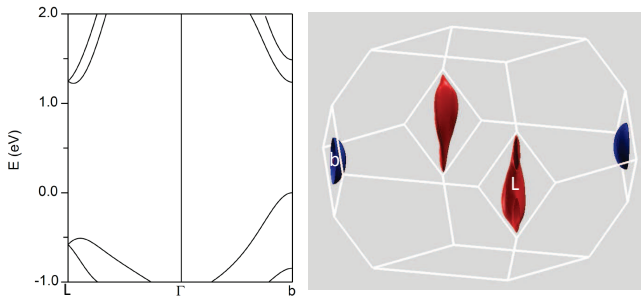}
\caption{Band structure (left) and carrier pockets (right). The point
b is (1/2,0,-1/2) in primitive cell reciprocal
lattice units and are shown in the zone in the right panel. The
carrier pockets are isosurfaces 0.05 eV below the valence band maximum (blue)
and 0.05 eV above the conduction band minimum (red). Note that the gap is
indirect.}
\label{fig-band}
\end{figure}

The band structure near the band edges is depicted in Fig. \ref{fig-band},
which shows the band structure along lines where the band extrema occur
and a isosurface visualization of the band edges. 
The valence band maximum (VBM) is on a zone face, as shown,
at the point denoted ``b".
The conduction band minimum (CBM) is near, but not at, another
zone face (L).
This indirect band gap has a value, $E_g(ind)$=1.22 eV.
The direct gap, $E_g(dir)$ is at ``b", and is only 0.02 eV
($\sim$250 K) larger.
This structure can provide a partial explanation for observed
good collection of photoexcited carriers.
Specifically, while the very small difference between $E_g(ind)$
and $E_g(dir)$ is insignificant from the point of view of obtaining
good optical absorption for the solar spectrum,
the indirect nature of the gap will impede recombination of
photoexcited carriers that relax to the band edges.
This effect will be stronger at room temperature if
the difference between the direct and indirect gaps is a little larger,
which is possible considering uncertainties in DFT calculations.
We note that 0.02 eV is a small energy and so it will  be of importance
to verify whether the gap is indirect and if so the magnitude of the
difference between the direct and indirect gaps by experiment.

\begin{figure}
\includegraphics[height=\columnwidth,angle=270]{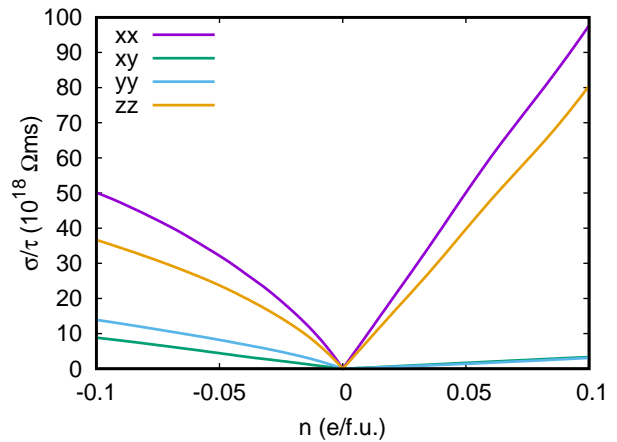}
\caption{Rigid band conductivity transport function $\sigma/\tau$, calculated
at 300 K, as a function of carrier concentration in electrons per
formula unit. Negative values denote holes.}
\label{fig-sigma}
\end{figure}

\begin{figure}
\includegraphics[height=0.9\columnwidth,angle=270]{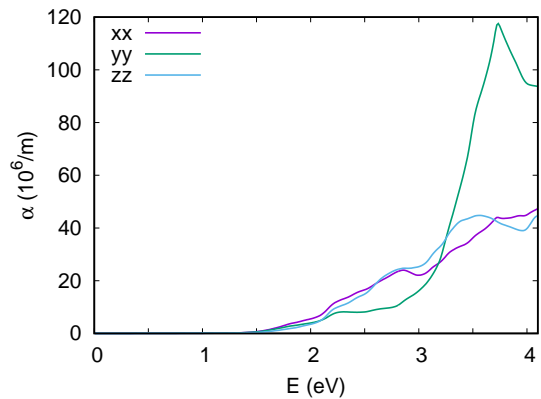}
\caption{Optical absorption spectrum. A Lorentzian broadening of 0.025 eV
was applied.}
\label{fig-abs}
\end{figure}

Fig. \ref{fig-sigma} shows the transport function $\sigma/\tau$ as obtained
from the electronic structure. As seen, the transport is highly two dimensional
for both the conduction and valance bands,
but is more so for the conduction bands. Transport in plane is also
anisotropic, with better conduction along the $a$ ($x$-direction) than
along $c$. This amounts to $\sim$35\% for the valence bands and $\sim$20\%
for the conduction bands. Finally, if the effective scattering rates,
$\tau^{-1}$, are similar for electrons and holes, the in-plane
mobility will be higher for electrons than for holes.

The calculated optical absorption spectrum is given in Fig. \ref{fig-abs}.
This spectrum was calculated in the independent particle approximation,
i.e. neglecting excitonic effects. These are anticipated to be
small due to the small band gap and resulting high electronic (clamped
ion) dielectric constant.
The spectrum is similar for both in plane polarizations but differs
strongly for the yy component, which has electric field polarization
perpendicular to the NaSbS$_2$ sheets.
Regardless of polarization, the absorption is relatively weak from the onset
at the direct gap to $\sim$ 2.5 eV.
This emphasizes the important role of good carrier transport to realize the
reported efficacy of this material as a solar absorber. \cite{rahayu}

\begin{figure}
\includegraphics[height=0.9\columnwidth,angle=270]{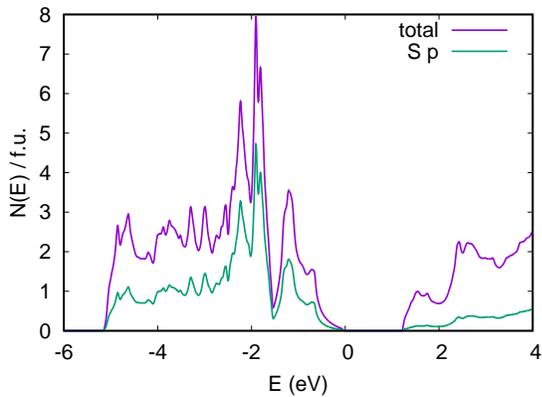}
\caption{Electronic density of states and S $p$ projection onto the LAPW
sphere on a per formula unit basis.}
\label{fig-dos}
\end{figure}

Fig. \ref{fig-dos} shows the calculated electronic density of states along with
the S $p$ contribution. The valence bands are derived from S $p$
states, so that the compound should be regarded as nominally ionic.
The top of the valence band manifold also appears to be split off to 
higher energy. This type of splitting is seen in some other S compounds where
it arises from repulsion between a lower lying metal state and the S $p$
states.
\cite{chen,mitzi}
The result is that the top of the valence band has antibonding
metal - S $p$ character, and often more dispersive bands beneficial for
transport.

\begin{figure}
\includegraphics[height=0.9\columnwidth,angle=270]{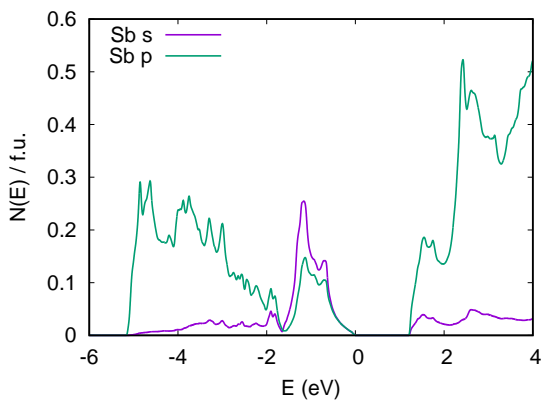}
\caption{Sb $s$ and $p$ density of states by projection onto the Sb LAPW
sphere. Note that the extended Sb valence orbitals lie mainly outside
the 2.2 bohr LAPW spheres,
so that the plot shows a quantity proportional to but
considerably smaller than the full Sb contributions.}
\label{fig-dos-sb}
\end{figure}

This antibonding mechanism is operative here. Fig. \ref{fig-dos-sb} shows
$s$ and $p$ projections on the Sb LAPW spheres. As seen, there is Sb $s$
character at the top of the valence bands including at the VBM. The main
Sb $s$ bands are at -9.5 to -7 eV relative to the VBM, and so the VBM
has S $p$ - Sb $5s$ antibonding character.

Besides this Sb $5s$ character at the top of the valence bands, there is
considerable Sb $5p$ character in the valence bands. The Sb $5p$ states
are nominally unoccupied in this compound and form the main conduction
bands. Thus this Sb $5p$ contribution in the
valence bands comes from cross-gap hybridization
between the occupied S $p$ states and unoccupied Sb $p$ states.
Such cross-gap
hybridization is a characteristic of oxide ferroelectric materials where it
leads to enhanced Born effective charges and thus ferroelectricity,
\cite{cohen}
and is also found in phase change materials.
\cite{mukhopadhyay}
It is closely connected with the concept of lone pair driven distortions.
Enhanced Born effective charges have also been associated with
efficient carrier transport in a number of materials.
\cite{du2,du1,fabini,brandt,lehner,du}
The mechanism is enhanced local screening due to the high Born charge,
which leads to defect tolerance in soft lattice materials. \cite{du2,brandt}

The Born effective charges were obtained
as ${\bf Z}_{ij}^{\ast}$ =
$\frac{\Omega}{e}\frac{\partial {\bf P}_{i}}{\partial {\bf u}_{j}}$,
where $\Omega$ is the volume of the unit cell,
${\bf P}_{i}$ is the total polarization in direction
$i$ and ${\bf u}_{j}$ is the displacement in direction $j$.
The calculated Born effective charges of monoclinic NaSbS$_{2}$
are shown in Table \ref{tab-mono-born}.
It can be seen that the maximum Born effective charges are
1.47 for Na, 4.69 for Sb and -3.08 for S,
respectively.
These are considerably larger than the corresponding nominal charges,
consistent with the expectation from the electronic structure.
The dielectric tensor contains both the electronic and ionic
contributions as
$\epsilon_{ij}$ = $\epsilon_{\infty,ij}$ + $\epsilon_{ph,ij}$.
The electronic part was obtained with ion-clamped using DFPT.
\cite{Gajdo,baroni}
The ionic contribution was based on the interatomic force constants
calculated using DFPT.\cite{wu}
For monoclinic NaSbS$_{2}$, there are four non-zero components,
as given in Table \ref{tab-mono-dielectric}.
The average value given by one third of the trace is 23.8.
For comparison, ZnO, which is a good oxide semiconductor that
has some defect tolerance at least for $n$-type,
\cite{look,mccluskey}
has a dielectric constant of 9.3.
\cite{crisler}

\begin{table}
\caption{Calculated Born effective charge tensors of monoclinic NaSbS$_2$.}
\begin{tabular}{lccccccccc}
\hline
   & $xx$ & $xy$ & $xz$ & $yx$ & $yy$ & $yz$ & $zx$ & $zy$ & $zz$ \\
\hline
Na & 1.13 & -0.15 & 0.00 & -0.15 & 1.47 & 0.00 & 0.00 & 0.00 & 1.19 \\
Sb & 3.43 &  1.55 & 0.00 &  1.51 & 4.69 & 0.00 & 0.00 & 0.00 & 2.31 \\
S  &-2.28 & -0.70 & 1.38 & -0.68 &-3.08 & 0.57 & 0.80 & 0.02 &-1.75 \\
\hline
\end{tabular}
\label{tab-mono-born}
\end{table}

\begin{table}
\caption{Calculated dielectric tensors of monoclinic NaSbS$_2$.}
\begin{tabular}{lcccc}
\hline
                       & $xx$ & $xy$  & $yy$ & $zz$ \\
\hline
$\epsilon_{\infty,ij}$ & 9.3  & 1.7   & 8.4  & 7.4  \\
$\epsilon_{ph,ij}$     &10.4  &13.0   &32.7  & 3.3  \\
$\epsilon_{ij}$        &19.7  &14.7   &41.1  &10.7  \\
\hline
\end{tabular}
\label{tab-mono-dielectric}
\end{table}

Therefore NaSbS$_2$ has enhanced Born effective charges due to
the cross gap hybridization, similar to several materials
that have been found to have excellent charge collection in the
context of radiation detection, e.g. TlBr, BiI$_3$ and Tl$_6$SeI$_4$.
\cite{du2,du1,biswas}
This enhanced Born charge leads to an enhanced dielectric constant,
which means enhanced screening.
This provides an explanation of how
a material that presumably contains high concentrations of point
defects can nonetheless have efficient carrier collection in an
optoelectronic application. In this regard, we note that SrTiO$_3$,
which is near a ferroelectric transition, and consequently has a very
high dielectric constant at low temperature, also has an exceptional
electron mobility that exceeds 30,000 cm$^2$/Vs in high quality films.
\cite{son}

\section{Summary and Conclusions}

We report first principles calculations for NaSbS$_{2}$.
We find that the ground state structure is monoclinic $C$2/$c$ or
possibly triclinic $P\bar{1}$ with a very small triclinic distortion.
We find highly anisotropic electronic
and optical properties as may be expected based on the crystal structure.
The results show a quasidirect band gap, with an indirect gap slightly
lower than the direct gap, which may impede carrier recombination.
The calculated value of the band gap using the mBJ potential is 1.22 eV.
Importantly, the electronic structure shows a substantial cross gap
hybridization between S $p$ and Sb $p$ states. This results in enhanced
Born effective charges and a high dielectric constant. This high
dielectric constant provides screening and defect tolerance for the
carrier collection. Therefore it is likely that NaSbS$_2$ can be
a useful optoelectronic material, not only as a solar absorber, but
also in applications requiring doping.

\acknowledgments

This work was supported by the Department of Energy through the MAGICS
Center, Award DE-SC0014607.

%

\end{document}